\documentclass{article}

\usepackage{amsmath, amssymb, xcolor, booktabs,framed}
\usepackage{algpseudocode}

\usepackage[utf8]{inputenc}
\usepackage{booktabs}
\usepackage{array}

\usepackage{ragged2e}
\usepackage[utf8]{inputenc}  
\usepackage{multirow}

\usepackage{PRIMEarxiv}
\usepackage{float}

\usepackage[utf8]{inputenc} 
\usepackage[T1]{fontenc}    
\usepackage{hyperref}       
\usepackage{url}            
\usepackage{booktabs}       
\usepackage{amsfonts}       
\usepackage{nicefrac}       
\usepackage{microtype}      
\usepackage{lipsum}
\usepackage{fancyhdr}       
\usepackage{graphicx}       
\graphicspath{{media/}}     

\title{AI-Driven IRM : Transforming Insider Risk Management with Adaptive Scoring and LLM-Based Threat Detection}

\author{
 Lokesh  Koli \\
Vectoredge\\
\texttt{lokesh@vectoredge.io} \\
   \And
 Shubham Kalra \\
Vectoredge\\
  \texttt{shubham@vectoredge.io} \\
  \And
Rohan Thakur \\
Vectoredge\\
  \texttt{rohan.t@vectoredge.io} \\
 \AND
Anas Saifi\\
Vectoredge\\
  \texttt{anas.s@vectoredge.io} \\
  \And
Karanpreet Singh \\
Vectoredge\\
  \texttt{karanpreet.s@vectoredge.io} \\
  \\
}

\begin{document}
\maketitle
\begin{abstract}
Insider threats pose a significant challenge to organizational security, often evading traditional rule-based detection systems due to their subtlety and contextual nature. This paper presents an AI-powered Insider Risk Management (IRM) system that integrates behavioral analytics, dynamic risk scoring, and real-time policy enforcement to detect and mitigate insider threats with high accuracy and adaptability. We introduce a hybrid scoring mechanism—transitioning from the static PRISM model to an adaptive AI-based model utilizing an autoencoder neural network trained on expert-annotated user activity data. Through iterative feedback loops and continuous learning, the system reduces false positives by 59\% and improves true positive detection rates by 30\%, demonstrating substantial gains in detection precision. Additionally, the platform scales efficiently, processing up to 10 million log events daily with sub-300ms query latency, and supports automated enforcement actions for policy violations, reducing manual intervention. The IRM system's deployment resulted in a 47\% reduction in incident response times, highlighting its operational impact. Future enhancements include integrating explainable AI, federated learning, graph-based anomaly detection, and alignment with Zero Trust principles to further elevate its adaptability, transparency, and compliance-readiness. This work establishes a scalable and proactive framework for mitigating emerging insider risks in both on-premises and hybrid environments.
\end{abstract}



\section{Introduction}
Insider threats arise when individuals with legitimate access to employees, contractors, or business partners misuse their privileges, either deliberately or inadvertently, leading to security breaches, data leaks, or operational disruptions. Unlike external cyber threats that originate from malicious actors outside an organization, insider threats exploit legitimate access privileges, making them more challenging to detect and mitigate.

As digital transformation accelerates, insider threats have grown more critical due to the rapid expansion of remote work, cloud adoption, and the surge in sensitive data storage. Organizations now depend heavily on cloud-based collaboration tools, remote work environments, and distributed identity systems, expanding the attack surface. Insiders can \textbf{leak sensitive data, manipulate records, or disrupt operations} using authorized credentials, often bypassing conventional security controls. According to various cybersecurity studies, insider threats are responsible for a substantial portion of \textbf{data breaches}\cite{verizon2024}, leading to \textbf{financial losses, reputational damage, and regulatory penalties}.\cite{ponemon2022}

Insider threats pose a significant challenge to cybersecurity, with three primary categories: \textbf{malicious, negligent, and accidental}. \textbf{Malicious insiders} intentionally misuse their access to harm an organization, steal sensitive data, or sabotage systems, such as employees leaking confidential files for financial gain or disgruntled staff deleting critical information. \textbf{Negligent insiders} expose organizations to risk through carelessness, like misconfiguring access permissions, using weak passwords, or disregarding security protocols. \textbf{Accidental insider threats} arise from unintentional mistakes, including sending sensitive documents to the wrong recipient, accidentally deleting data, or falling victim to phishing attacks.

Detecting and mitigating these threats is complex due to the challenge of differentiating \textbf{legitimate user activities from subtle malicious behaviors}. \textbf{Traditional rule-based systems} struggle with \textbf{dynamic access patterns}\cite{Malik2024}, often leading to \textbf{high false positives or undetected risks}. A more effective approach involves \textbf{AI-driven risk assessment}, continuously analyzing \textbf{user actions, access patterns, and deviations from normal behavior}. Insiders can \textbf{exfiltrate data} through \textbf{encrypted channels, cloud storage, or personal devices}, making \textbf{real-time visibility and forensic tracking} essential. However, \textbf{balancing security and privacy} remains crucial, as \textbf{excessive monitoring} raises \textbf{legal and ethical concerns}\cite{nist_privacy}, while \textbf{inadequate oversight} creates \textbf{security blind spots}.

Our \textbf{AI-enabled} Insider Risk Management (IRM) system integrates \textit{PRISM– Privilege-based Risk \& Insider Scoring Mechanism}, AI-enabled risk scoring, behavior-based anomaly detection\cite{ponemon2022}, Policy-Based Risk Analysis \& Response Automation, and context-aware alerts to address these gaps. This system offers \textbf{real-time behavioral analytics, dynamic risk scoring, and automated mitigation}\cite{gartner2023}, enabling organizations to \textbf{avoid insider threats}. Its \textbf{multi-platform connectivity} ensures \textbf{seamless integration} with \textbf{identity providers (IDPs)} like \textbf{Azure AD\cite{azuread}, AWS IAM\cite{awsiam}, and Google Workspace\cite{googleworkspace}}, alongside \textbf{enterprise applications} such as \textbf{SharePoint, OneDrive, Microsoft Teams, Box, Slack, Salesforce, Google Workspace, etc.}

The \textbf{AI-driven risk scoring model and PRISM} continuously evaluate \textbf{user activities} across various categories, including:

\begin{itemize}
    \item \textbf{User Risk} – Detects \textbf{login anomalies, credential misuse, and unauthorized access attempts}.
    \item \textbf{Data Movement Risk} – Identifies \textbf{covert file transfers, suspicious downloads, and improper document sharing}.
    \item \textbf{Attack Path Risk} – Maps \textbf{vulnerabilities in infrastructure using knowledge graphs}.
    \item \textbf{Activity Risk} – Monitors \textbf{unusual logins, device access, and location-based anomalies}.
    \item \textbf{Data Risk} – Tracks \textbf{sensitive data access, deletion, and storage policies}.
    \item \textbf{Data Collaboration Risk} – Prevents \textbf{unauthorized sharing of sensitive documents}.
\end{itemize}

Leveraging \textbf{large language models (LLMs)}\cite{openai2023}, the system dynamically \textbf{analyzes risk scores and user behavior}. It enhances contextual understanding to generate AI-generated actionable recommendations, which analyze the issues and suggest appropriate actions for the Security Expert.

Moreover, \textbf{security teams} gain access to \textbf{interactive dashboards} delivering \textbf{real-time risk scores, behavioral anomalies, and system activity insights}, allowing \textbf{proactive threat mitigation}\cite{gartner2023}. As \textbf{insider threats evolve}, \textbf{traditional security measures fail to provide} \textbf{real-time intelligence} for \textbf{effective detection and response}\cite{nist800-53}. Organizations must transition from \textbf{reactive security strategies} to \textbf{AI-driven, proactive risk management}.

By leveraging \textbf{behavioral analysis\cite{ponemon2022}, Prism, AI-based risk assessment, and context-aware Recommendations}, businesses can \textbf{enhance threat detection, ensure compliance, minimize false positives, and strengthen their cybersecurity posture}. In this \textbf{evolving threat landscape}, \textbf{AI-powered risk assessment} positions organizations \textbf{ahead of potential security risks}, safeguarding their \textbf{most valuable assets}.

\section{Background and Related Work}
\label{sec:headings}
Insider risk management has traditionally relied on rule-based security models, manual audits, and behavior monitoring tools\cite{nist800-53}. Conventional approaches focus on access controls, user activity logs, and predefined security policies to detect unauthorized behavior. Security Information and Event Management (SIEM) systems, User and Entity Behavior Analytics (UEBA), and Data Loss Prevention (DLP) tools have been widely used in enterprises to monitor suspicious activities\cite{gartner2023}. However, these solutions often generate high volumes of alerts, many of which are false positives, making it challenging for security teams to prioritize real threats\cite{ponemon2022}.

Traditional insider risk management approaches primarily depend on \textbf{static rules} and \textbf{signature-based detection}, which fail to adapt to evolving insider threats. These methods cannot detect \textbf{subtle, context-dependent behaviors}, such as \textbf{progressive data exfiltration, privilege misuse, or slow insider reconnaissance}\cite{verizon2023}.

AI-enabled insider risk management, on the other hand, leverages \textbf{machine learning (ML), natural language processing (NLP), and behavioral analytics} to analyze vast amounts of user activity data in real-time\cite{openai2023}. By continuously learning from user behavior patterns, AI-driven models can detect \textbf{anomalous activities}, assess \textbf{contextual risks}, and \textbf{generate adaptive risk scores}\cite{gartner2023ai}.

\begin{table}[H]
\centering
\renewcommand{\arraystretch}{1.3}
\caption{Comparison of Traditional Methods vs. AI-Driven Approaches}
\label{tab:traditional_vs_ai}
\begin{tabular}{|p{4cm}|p{5cm}|p{5cm}|}
\hline
\textbf{Feature} & \textbf{Traditional Methods} & \textbf{AI-Driven Approaches} \\
\hline
\textbf{Detection Mechanism} & Rule-based and manual thresholds & Behavioral analytics and anomaly detection \\
\hline
\textbf{Adaptability} & Static, predefined rules & Dynamic, continuously learning models \\
\hline
\textbf{False Positives} & High, due to lack of contextual analysis & Reduced, with contextual and behavioral insights \\
\hline
\textbf{Data Processing} & Limited historical analysis & Real-time, large-scale data processing \\
\hline
\textbf{Risk Scoring} & Basic, manual assessment & Automated, AI-enhanced scoring \\
\hline
\textbf{Response and Recommendations} & Reactive, requiring manual intervention & Proactive, automated recommendations \\
\hline
\end{tabular}
\end{table}

Despite advancements in AI-based security tools, several critical gaps remain in existing insider risk management solutions:

\begin{enumerate}
    \item \textbf{Lack of Real-Time Risk Scoring}—Many existing solutions rely on periodic log analysis rather than real-time monitoring, which leads to delayed detection and response\cite{gartner2023}.
    \item \textbf{Incomplete Data Lineage Tracking}—Most traditional systems struggle to track\textbf{ the entire lifecycle of sensitive data}, from creation and modification to sharing and deletion, particularly across \textbf{hybrid and multi-cloud environments}\cite{ponemon2022}.
    \item \textbf{High False Positives} – A significant challenge in insider risk detection is \textbf{distinguishing between legitimate activity and genuine threats}, as static rule-based systems generate excessive false alarms\cite{verizon2023}.
    \item \textbf{Limited Integration with Modern Workflows} – Many security tools do not seamlessly integrate with \textbf{collaboration platforms, cloud services, and hybrid infrastructure}, leading to blind spots in insider risk monitoring\cite{gartner2023ai}.
\end{enumerate}

These limitations highlight the need for an \textbf{AI-driven risk management system} incorporating \textbf{real-time anomaly detection, automated risk scoring, and data lineage tracking} across various platforms\cite{openai2023}.

\section{\textbf{Materials and methods}}
\label{sec:others}
\subsection{Dataset}
The CERT Insider Threat Dataset is a widely recognized benchmark dataset designed for studying insider threats within organizations. Developed by the Carnegie Mellon University Software Engineering Institute (CMU-SEI), it simulates real-world enterprise environments by generating synthetic yet realistic user activity logs \cite{glasser2013bridging, certdataset2016}. The dataset includes multiple log sources such as authentication records, file accesses, email communications, and psychometric assessments, providing a comprehensive foundation for insider threat detection research \cite{salem2008survey}. Its primary advantage lies in capturing both benign and malicious insider activities, allowing for developing and evaluating advanced security analytics and AI-driven risk-scoring models \cite{greitzer2013combining}.

To develop an AI-driven insider risk scoring model, we utilized the CERT dataset, incorporating multiple log sources such as user activity (users.csv), authentication records (logon.csv), file access events (file.csv) and device interactions (device.csv) \cite{certdataset2016}. Given the structured nature of these logs, we first preprocessed the dataset by filtering out irrelevant columns, retaining only the parameters relevant to our insider risk framework. To ensure high-quality annotations, we leveraged domain expertise from security professionals to analyze user behavior, assign appropriate risk scores, and validate threat classifications [internal methodology – may not require citation unless using external standards].

These annotated datasets were then processed through our \textit{PRISM – Privilege-based Risk \& Insider Scoring Mechanism,} which assesses insider risk based on predefined security metrics and behavioral patterns. The performance of this \textit{PRISM} serves as a baseline for comparison with our AI-driven risk-scoring approach, which we will discuss in upcoming sessions.

Additionally, the dataset underwent further preprocessing steps, including normalization, timestamp alignment, and event correlation, to closely resemble real-world data collected from our enterprise security pipeline [custom process – not publicly citable]. This enriched dataset laid the foundation for training our initial AI-based risk-scoring model. Finally, we integrated real-time data streams from our production environment with the CERT dataset to enhance model generalization, ensuring that our system adapts dynamically to emerging insider threat patterns [practical engineering – internal claim].

\subsection{System Architecture}
\begin{figure}[H]
    \centering
    \includegraphics[width=0.5\linewidth]{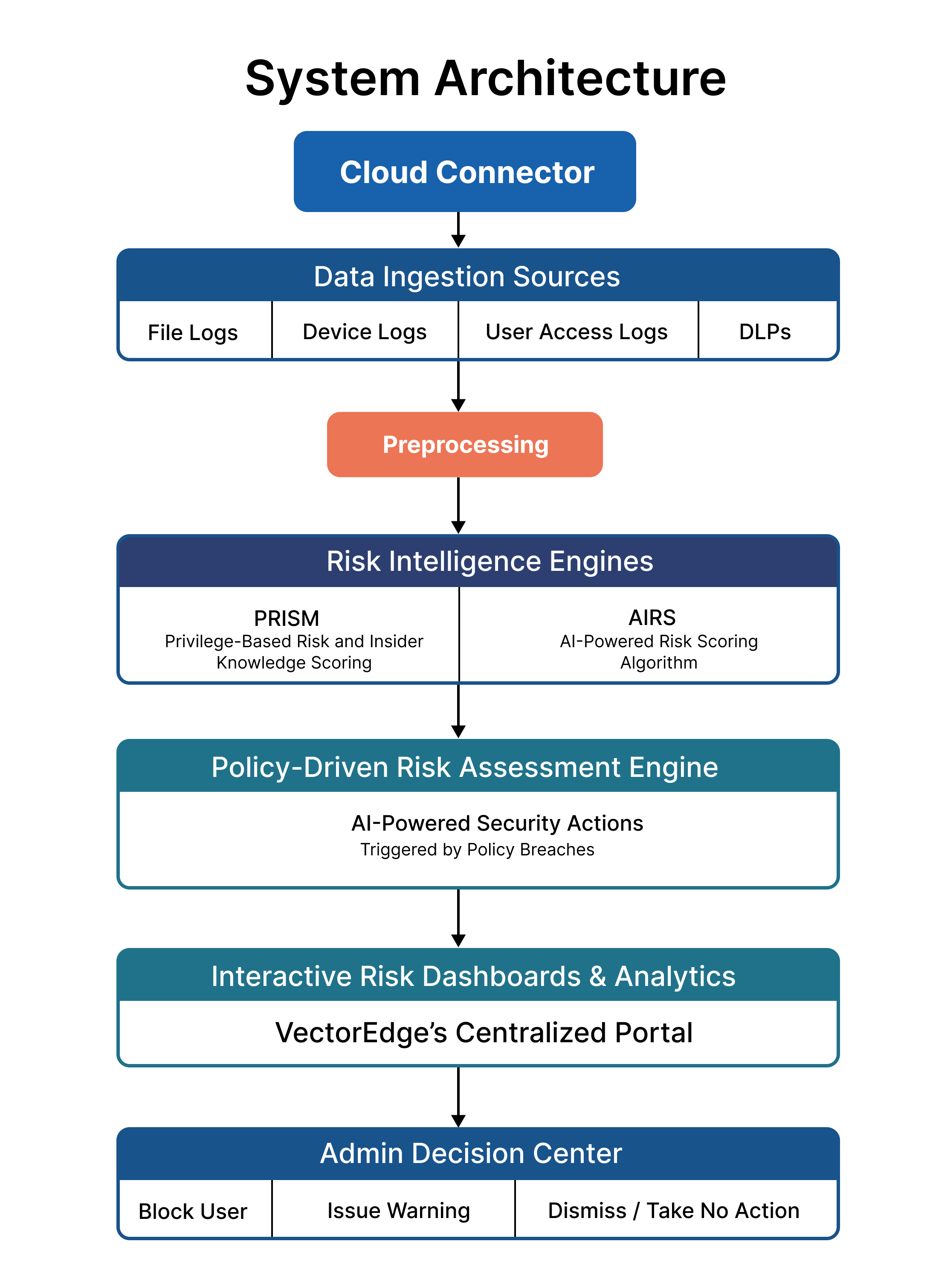}
    \caption{System architecture of the proposed AI-enabled insider risk management framework.}
    \label{fig:system_architecture}
\end{figure}

The proposed AI-enabled insider risk management system is designed to efficiently process and analyze diverse security logs, enabling the detection and mitigation of insider threats. It integrates multiple data sources, utilizes PRISM and AI-driven risk scoring, conducts anomaly detection, and enforces policy-based risk assessments to generate actionable security insights, as illustrated in Figure~\ref{fig:system_architecture}.

\subsubsection{Data Sources and Collection}
The system continuously ingests logs and activity data from multiple sources across an organization’s IT infrastructure to provide a comprehensive and real-time security assessment. These diverse data points enable detailed risk analysis, insider threat detection, and compliance monitoring.

\begin{enumerate}
    \item Device Logs
    \begin{itemize}
        \item Captures system-level activity, including processes executed, applications accessed, and system interactions.
        \item It helps identify unauthorized device usage, unusual locations, and other operating system-based suspicious activities.
    \end{itemize}
    \item User Activity Logs
    \begin{itemize}
        \item Tracks user actions across enterprise applications, collaboration platforms, and cloud services.
        \item Logs capture file modifications, data transfers, messaging activities in applications like Teams, and administrative changes.
        \item Provides insights into policy violations, anomalous behavior, and insider risks.
    \end{itemize}
    \item Login Logs
    \begin{itemize}
        \item Monitors all authentication events, including successful and failed login attempts.
        \item Detects brute force attacks, unauthorized access attempts, and login anomalies such as location-based inconsistencies.
    \end{itemize}
    \item File Access Logs
    \begin{itemize}
        \item Records user interactions with files, including reading, writing, deleting, and sharing actions.
        \item Identifies sensitive data movement, potential exfiltration attempts, and unauthorized access to critical files.
    \end{itemize}
\end{enumerate}
One of the system's core capabilities is its ability to analyze real-time file sensitivity. The cutting-edge Context-Aware Hybrid Pattern Detection Algorithm (CHPDA) is leveraged to classify files based on their content and compliance risk factors~\cite{chpda2025}. This AI-supported detection mechanism identifies and categorizes files containing Personally Identifiable Information (PII), Protected Health Information (PHI), Payment Financial Information (PFI), and other compliance-regulated data (e.g., GDPR, HIPAA, PCI-DSS).

\begin{enumerate}
    \item Sensitive File Classification:
    \begin{enumerate}
        \item Text-Based Analysis – Scans file content to detect sensitive information.
        \item Metadata \& Contextual Analysis – Examines file names, locations, and historical access patterns.
        \item Behavior-Based Classification – Identifies suspicious data sharing, mass file deletions, and unauthorized transfers.
    \end{enumerate}
    \item Automated Compliance Mapping:
    \begin{enumerate}
        \item Ensures that all files comply with enterprise security policies and industry regulations.
        \item Flags files that violate data handling policies or are at risk of unauthorized exposure.
    \end{enumerate}
\end{enumerate}
Logs and activity data are collected through seamless connectors that integrate with various enterprise applications, cloud services, and on-premise environments:

\begin{enumerate}
    \item Identity Providers \& Authentication Systems
Azure AD, AWS IAM, Google Cloud Identity – Tracks user identity and access activities.

    \item Enterprise Collaboration \& Cloud Storage Platforms
SharePoint, OneDrive, Google Drive, Microsoft Teams, Box – Monitors file sharing, access control, and data movement.

    \item On-Premises \& Hybrid IT Infrastructure
Windows file shares, local IAM systems, hybrid cloud setups – Captures data from traditional enterprise environments.

\end{enumerate}
All collected logs are securely stored in an on-premises, encrypted database. The data then undergoes preprocessing, which includes normalization and standardization to ensure consistency across different log sources, as well as timestamp synchronization to align logs from multiple systems for accurate event correlation. This structured dataset serves as the backbone for real-time risk scoring, anomaly detection, and automated security response mechanisms.

\subsubsection{PRISM – Privilege-based Risk \& Insider Scoring Mechanism}
Once the data is collected from various sources, the system applies a \textbf{PRISM} – \textbf{Privilege-based Risk \& Insider Scoring Mechanism} that assigns a risk score R based on predefined rules and security heuristics~\cite{liu2018risk,greitzer2012predicting,salem2008survey}. The algorithm evaluates user activity by considering multiple parameters, each contributing weight to the final risk score. The total risk score is computed as follows:

\begin{equation}
R = (W_p \cdot S_p) + (W_A \cdot S_A) + (W_C \cdot S_C) + (W_{IP} \cdot S_{IP}) + (W_B \cdot S_B) + (W_D \cdot S_D) + (W_{CA} \cdot S_{CA})
\end{equation}

where:

\begin{itemize}
\item $S_P$ = \textbf{User Privilege Score}
\item $S_A$ = \textbf{Activity Type Score}
\item $S_C$ = \textbf{Application Context Score}
\item $S_{IP}$ = \textbf{IP Address Score}
\item $S_B$ = \textbf{Business Hours Score}
\item $S_D$ = \textbf{Device Compliance Score}
\item $S_{CA}$ = \textbf{Cumulative Activity Score}
\item $W_x$ represents the \textbf{weight assigned to each factor}

\end{itemize}

\textit{\textbf{Risk Scoring Matrix}}

\textbf{Example Calculation Scenario:}

A low-privilege employee logs in from an unknown IP address, accesses SharePoint, moves five files, and performs these actions outside business hours from a non-compliant device.

For simplicity, let us assume the base risk score \(R_0=0\), and all weights are equally set at \(W=1\).

\begin{table}[H]
    \caption{Risk Score Impact Based on Different Factors}
    \centering
    \begin{tabular}{llc}
        \toprule
        \textbf{Factor} & \textbf{Condition} & \textbf{Impact on Risk Score (S)} \\
        \midrule
        \multirow{3}{*}{User Privileges ($S_P$)} 
        & High-privilege admin roles & $R \times 0.5 - 0.9$ (decrease) \\
        & Moderate-privilege roles & $R \times 0.8 - 0.95$ (decrease) \\
        & Low-privilege roles/guests & $R \times 1.1$ (increase) \\
        \midrule
        \multirow{6}{*}{Activity Type ($S_A$)}
        & File upload (low impact) & +1 \\
        & File creation & +2 \\
        & Attachment shared/edited & +3 \\
        & File rename/move & +4 - 5 \\
        & File shared externally & +7 \\
        & File deletion (high impact) & +8 \\
        \midrule
        Application Context ($S_C$) & OneDrive, SharePoint, Teams & Context-dependent risk \\
        \midrule
        \multirow{2}{*}{IP Address Reputation ($S_{IP}$)} 
        & Known \& trusted IP & No impact \\
        & Unknown or blacklisted IP & +5 \\
        \midrule
        \multirow{2}{*}{Business Hours ($S_B$)} 
        & Activity between 9 AM - 5 PM & No impact \\
        & Activity outside business hours & +5 \\
        \midrule
        \multirow{2}{*}{Device Compliance ($S_D$)} 
        & A managed and compliant device & No impact \\
        & Unmanaged / non-compliant device & +5 (each) \\
        \midrule
        Cumulative Activity ($S_{CA}$) & Excessive/repetitive actions & Progressive risk increase \\
        \bottomrule
    \end{tabular}
    \label{tab:risk_score_impact}
\end{table}

\textit{\textbf{Step 1: Calculate risk contributions}}

To begin the risk scoring process, we analyze individual contributing factors based on the user's behavior and context, as detailed in Table~\ref{tab:risk_score_impact}. The user has a \textbf{low-privilege role}, which does not directly add to the base score but introduces a \textbf{risk multiplier of 1.1}, applied in the final step. The user \textbf{moved five files}, each move contributing \textbf{4 points}, totaling \textbf{20 points} for activity type. The activity occurred on \textbf{SharePoint}, which in this context carries \textbf{no additional risk}. Logging in from an \textbf{unknown IP address} adds \textbf{+5 points}, performing the action \textbf{ outside of business hours} contributes another \textbf{+5 points}, and using a \textbf{non-compliant device} adds another \textbf{ + 5 points}.

Together, these individual components lead to a base risk score of \textbf{35}, which is later adjusted by the privilege multiplier to produce the final score, as outlined in Table~\ref{tab:risk_score_impact}.

\textbf{Step 2: Calculate the Total Score
}
\begin{equation}
R = (1 \times 20) + (1 \times 5) + (1 \times 5) + (1 \times 5) = 35
\end{equation}

Applying the low-privilege multiplier:

\begin{equation}
R = 35 \times 1.1 = 38.5
\end{equation}

\textbf{Step 3: Normalize Risk Score (0-1 Scale)}

To normalize the risk score within a range of 0-1, we apply Min-Max normalization:

\begin{equation}
R_{\text{norm}} = \frac{R - R_{\text{min}}}{R_{\text{max}} - R_{\text{min}}}
\end{equation}

Assuming the minimum risk score is 0, and the maximum possible risk score is 100, the normalized risk score is the following:

\begin{equation}
R_{\text{norm}} = \frac{38.5 - 0}{100 - 0} = 0.385
\end{equation}

\textit{Interpretation:}The final normalized risk score for this session is 0.385, classifying this activity as moderate risk based on the following thresholds:

\begin{itemize}
    \item 0.0 - 0.3 → Low Risk
    \item 0.3 - 0.6 → Moderate Risk
    \item 0.6 - 1.0 → High Risk
\end{itemize}
Since the risk threshold for a security alert is typically 0.3, this action would trigger an investigation. The security team would now assess whether this is a legitimate activity or an insider threat.

This approach ensures a structured and explainable risk assessment model, balancing static heuristics with contextual analysis to detect potential insider threats dynamically.

\subsubsection{ AIRS - AI Risk Scoring Algorithm: AI-Based Risk Scoring Framework}

The system employs an AI-driven risk-scoring model based on an autoencoder neural network to enhance the accuracy of insider threat detection. This approach improves upon traditional risk assessment methods by learning from historical data and adapting to evolving threats \cite{pantelidis2021insider}.

The AI model operates in the following stages:

\begin{enumerate}
    \item \textbf{Initial Training Phase}
    \begin{itemize}
        \item The AI model is initially trained using data from the PRISM framework.
        \item This data serves as labeled input, enabling the model to understand predefined risk patterns and behaviors.
        \item The autoencoder learns a baseline representation of normal and risky activities by identifying patterns in past user behavior.
        \item The system calculates a reconstruction error to measure deviations from routine behavior \cite{pantelidis2021insider}.
    \end{itemize}

    \item \textbf{User Feedback Loop}
    \begin{itemize}
        \item Once trained, the AI assigns a risk score to new activities based on their deviation from established patterns:
        \[
        S_{\text{AI}} = \text{normalize}(\text{Reconstruction Error})
        \]
        Higher reconstruction errors correspond to higher risk scores, scaled between 0 and 1 for consistency.
        \item Security analysts review the assigned risk scores to determine alignment with security expectations.
        \item Analysts can provide feedback to refine the AI model's assessments \cite{bentley2024humans}.
    \end{itemize}

    \item \textbf{Risk Score Adjustment via User Input}
    \begin{itemize}
        \item The system allows manual risk score adjustments through a slider-based interface to ensure flexibility:
        \[
        S_{\text{final}} = S_{\text{AI}} + \alpha (S_{\text{user}} - S_{\text{AI}})
        \]
        Where:
        \begin{itemize}
            \item $S_{\text{user}}$ = Analyst’s adjusted score
            \item $\alpha$ = A factor controlling the influence of user feedback
        \end{itemize}
        \item This feedback mechanism fine-tunes the AI model’s interpretation of risk factors.
    \end{itemize}

    \item \textbf{Incremental Model Relearning}
    \begin{itemize}
        \item The system maintains a threshold for retraining; once a set number of feedback instances are collected, the AI model undergoes incremental retraining.
        \item User feedback is prioritized over initial training weights during this process, ensuring the model adapts to the organization's unique risk patterns and security policies \cite{bentley2024humans}.
    \end{itemize}

    \item \textbf{Personalized Risk Profiling \& Continuous Learning}
    \begin{itemize}
        \item Over time, the AI model learns from past user inputs, improving its ability to distinguish between normal and high-risk activities.
        \item This continuous learning reduces false positives and enhances real-time risk assessment accuracy.
        \item The model evolves to align risk assessments with security operations rather than relying on rigid predefined rules.
    \end{itemize}

\textbf{Why Does This Approach Matter?} Integrating human-in-the-loop learning into the AI-based risk-scoring model ensures a dynamic, adaptable, and highly accurate security framework. Unlike static rule-based systems, this approach learns from security analysts, adapts to real-world threats, and continuously improves to provide more precise risk assessments, as illustrated in Figure~\ref{fig:hilt_architecture}.
 \cite{bentley2024humans}.
 
\begin{figure}[H]
    \centering
    \includegraphics[width=1\linewidth]{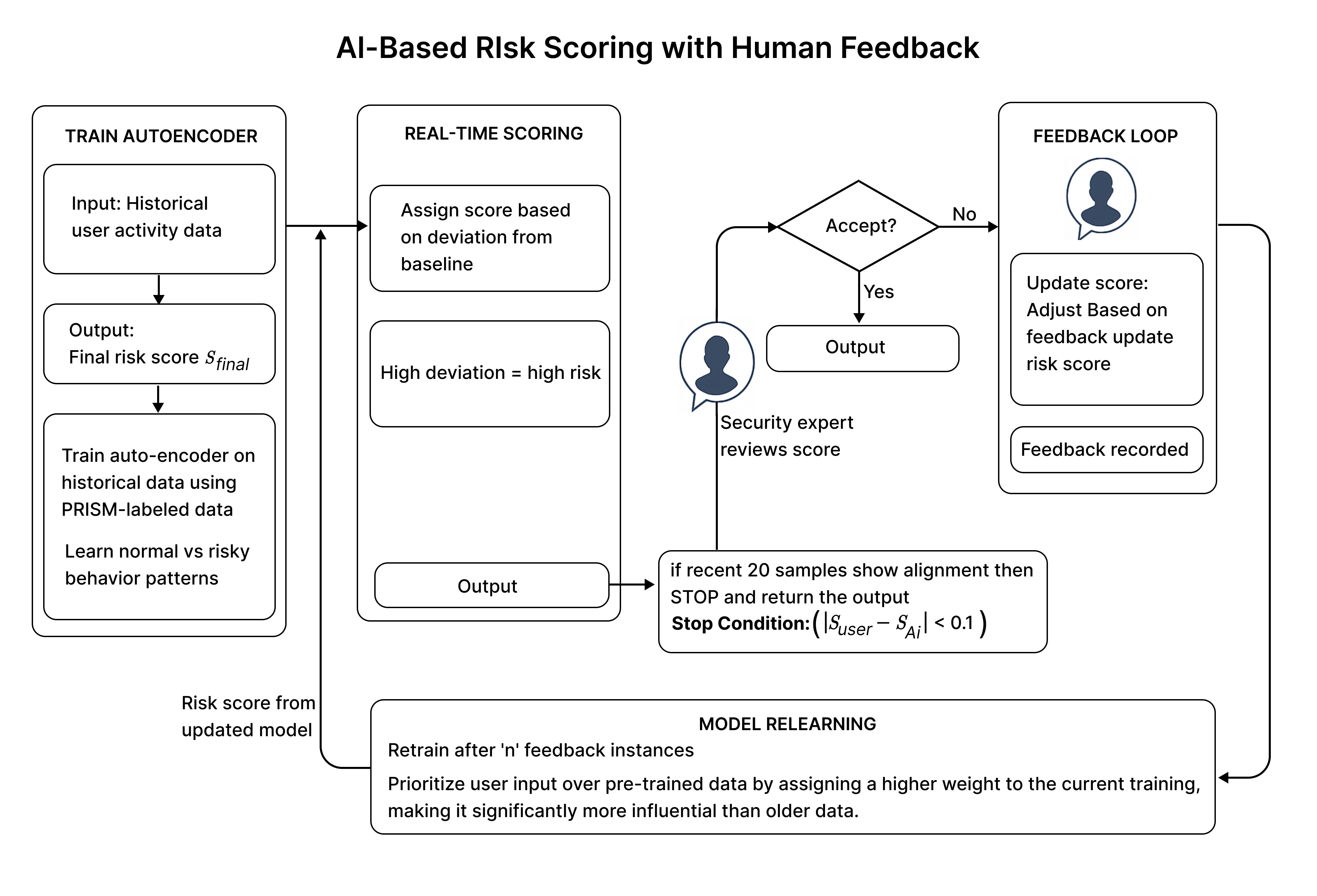}
  \caption{Human-in-the-loop learning architecture integrated with AI-based risk scoring.}
\label{fig:hilt_architecture}
\end{figure}

\begin{framed}
\textbf{AI-Driven Risk Scoring with Feedback Integration}

\begin{algorithmic}[1]
\State \textbf{Input:} Historical user activity data
\State \textbf{Output:} Final risk score $S_{\text{final}}$

\Procedure{Train Autoencoder}{}
    \State Train autoencoder on pretrained-prism data
    \State Learn standard user behavior patterns
\EndProcedure

\Procedure{RealTimeRiskAssessment}{activity}
    \State Compute reconstruction error for activity
    \State Normalize the error $\rightarrow S_{\text{AI}}$
\EndProcedure

\Procedure{IncorporateUserFeedback}{$S_{\text{AI}}$}
    \State Present $S_{\text{AI}}$ to analyst
    \If{analyst adjusts score}
        \State Update score: $S_{\text{final}} = S_{\text{AI}} + \alpha (S_{\text{user}} - S_{\text{AI}})$
    \Else
        \State $S_{\text{final}} = S_{\text{AI}}$
    \EndIf
\EndProcedure

\Procedure{UpdateModel}{}
    \If{user feedback collected for n activities}
        \State Incrementally retrain autoencoder with feedback
    \EndIf
\EndProcedure

\Procedure{ContinuousLearning}{}
    \State Update user’s cumulative risk profile
    \If{cumulative risk $>$ threshold}
        \State Trigger alert for investigation
        \If{Security Expert rejects the alert decision}
            \State Re-evaluate using \textsc{IncorporateUserFeedback}
            \State Retrain using \textsc{UpdateModel}
        \EndIf
    \EndIf
\EndProcedure
\end{algorithmic}

\end{framed}

\end{enumerate}

\subsubsection{Policy-Based Risk Analysis \& Response Automation}

The system enforces predefined security policies for various cloud storage and collaboration platforms like SharePoint, OneDrive, Google Drive, Teams, and Box. These policies act as the first line of defense by identifying and mitigating risks in real-time. When a security violation is detected, an alert is generated, ensuring that unauthorized or risky actions do not go unnoticed. These policies are categorized based on risk dimensions, providing comprehensive coverage across users, data, activity, and attack paths~\cite{microsoft_defender, uptycs_attack_path}.

\begin{enumerate}
    \item User Risk Policies
These policies detect anomalous user behaviors, including unauthorized access attempts, login anomalies, and potential credential misuse.
See Table ~\ref{tab:policy_risks}
\begin{table}[H]
    \centering
    \caption{Policy-Based Detection Triggers and Associated Risks}
    \label{tab:policy_risks}
    \begin{tabular}{>{\raggedright\arraybackslash}p{4.5cm} 
                    >{\raggedright\arraybackslash}p{6.5cm} 
                    >{\raggedright\arraybackslash}p{4.5cm}}
        \toprule
        \textbf{Policy Name} & \textbf{Trigger Condition} & \textbf{Potential Risk} \\
        \midrule
        Suspicious Login Activity & Multiple failed login attempts within a short time & Brute-force attack, credential stuffing \\
        Unusual Location Access & Login from a new/unapproved geographic region & Compromised credentials, unauthorized access \\
        Privileged Escalation Without Authorization & Sudden admin privilege grant & Insider threat, unauthorized control gain \\
        Login from Untrusted Device & Access from a device not previously associated with the user & Stolen credentials, unauthorized access \\
        \bottomrule
    \end{tabular}
\end{table}

\item Data Movement Risk Policies

These policies monitor unauthorized data transfers across cloud platforms to prevent data ex-filtration and information leaks as mention in Table \ref{tab:policy_triggers_risks}.

\begin{table}[H]
    \centering
    \caption{Policy Triggers and Associated Risks}
    \label{tab:policy_triggers_risks}
    \begin{tabular}{>{\raggedright\arraybackslash}p{4.8cm} 
                    >{\raggedright\arraybackslash}p{6.2cm} 
                    >{\raggedright\arraybackslash}p{4.5cm}}
        \toprule
        \textbf{Policy Name} & \textbf{Trigger Condition} & \textbf{Potential Risk} \\
        \midrule
        Mass Data Download & Unusually high file downloads from OneDrive/Google Drive & Insider threat, data theft \\
        External Sharing of Sensitive Data & File shared with an external, untrusted email domain & Data leakage, regulatory non-compliance \\
        Unapproved Cloud Sync & Data synced to unauthorized third-party storage & Shadow IT, unauthorized data transfer \\
        Unencrypted File Transfers & PII/PHI transferred without encryption & Compliance violations, data breach risk \\
        \bottomrule
    \end{tabular}
\end{table}

\item Attack Path Risk Policies

 Policies in Table\ref{tab:advanced_policy_risks} utilize knowledge graphs to map potential attack paths within an organization's infrastructure, identifying weak points before exploitation.

\begin{table}[H]
    \centering
    \caption{Advanced Risk Policy Triggers and Potential Threats}
    \label{tab:advanced_policy_risks}
    \begin{tabular}{>{\raggedright\arraybackslash}p{5.2cm} 
                    >{\raggedright\arraybackslash}p{5.6cm} 
                    >{\raggedright\arraybackslash}p{4.2cm}}
        \toprule
        \textbf{Policy Name} & \textbf{Trigger Condition} & \textbf{Potential Risk} \\
        \midrule
        Open Attack Paths via Misconfigured Access Controls & The user has excessive privileges in multiple systems & Lateral movement, privilege escalation \\
        Unpatched Vulnerability Exploitation & System/service running outdated software & Exploitable attack surface \\
        Multiple High-Risk Access Points & High-privilege user accesses multiple high-risk resources & Advanced persistent threat (APT) behavior \\
        \bottomrule
    \end{tabular}
\end{table}

\item Activity Risk Policies

These policies detect anomalies in user behavior by identifying suspicious logins, device access, and location-based risk factors Table\ref{tab:behavioral_anomalies}

\begin{table}[H]
    \centering
    \caption{Behavioral Anomaly Detection Policies}
    \label{tab:behavioral_anomalies}
    \begin{tabular}{>{\raggedright\arraybackslash}p{5cm} 
                    >{\raggedright\arraybackslash}p{6cm} 
                    >{\raggedright\arraybackslash}p{4cm}}
        \toprule
        \textbf{Policy Name} & \textbf{Trigger Condition} & \textbf{Potential Risk} \\
        \midrule
        Excessive Login Failures & Multiple failed login attempts from different devices & Credential stuffing, brute-force attacks \\
        Simultaneous Logins from Multiple Locations & Users log in from geographically distant locations within a short timeframe & Account compromise \\
        Unusual Device Access & A user accesses corporate systems from an unknown or unauthorized device & Stolen credentials, unauthorized access \\
        \bottomrule
    \end{tabular}
\end{table}

\item Data Risk Policies

These policies protect sensitive data assets by enforcing access restrictions, retention policies, and storage security.See Table \ref{tab:data_security_policies}

\begin{table}[H]
    \centering
    \caption{Data Security and Compliance Policy Triggers}
    \label{tab:data_security_policies}
    \begin{tabular}{>{\raggedright\arraybackslash}p{5cm} 
                    >{\raggedright\arraybackslash}p{6cm} 
                    >{\raggedright\arraybackslash}p{4cm}}
        \toprule
        \textbf{Policy Name} & \textbf{Trigger Condition} & \textbf{Potential Risk} \\
        \midrule
        Unauthorized Access to Sensitive Data & User attempts to access restricted PII/PHI & Insider threat, regulatory violation \\
        Mass Deletion of Critical Files & Bulk deletion of critical business files & Accidental data loss, ransomware attack \\
        Storage Policy Violation & Sensitive data stored in an unapproved location & Non-compliance, increased breach risk \\
        \bottomrule
    \end{tabular}
\end{table}

\item Data Collaboration Risk Policies

These policies focus on ensuring secure collaboration by preventing unauthorized document sharing and controlling information flow across teams and external entities. See Table \ref{tab:data_collab_policies}

\begin{table}[H]
    \centering
    \caption{Data Collaboration Risk Policies}
    \label{tab:data_collab_policies}
    \begin{tabular}{>{\raggedright\arraybackslash}p{5cm} 
                    >{\raggedright\arraybackslash}p{6cm} 
                    >{\raggedright\arraybackslash}p{4cm}}
        \toprule
        \textbf{Policy Name} & \textbf{Trigger Condition} & \textbf{Potential Risk} \\
        \midrule
        Sharing of Confidential Documents with External Parties & Sensitive documents shared outside the organization & Data leakage, regulatory non-compliance \\
        Public File Link Creation for Internal Data & A confidential document is shared via a public link & Unauthorized access, information leaks \\
        Abnormal Collaboration Behavior & A user shares a high number of files in a short period & Insider risk, data exfiltration \\
        \bottomrule
    \end{tabular}
\end{table}

\item Behavior-Based Anomalies

Rather than focusing on specific actions, these policies detect \textbf{unusual behavioral patterns} by analyzing deviations from normal usage. Examples include:
\begin{itemize}

\item \textbf{Excessive File Downloads:} A user downloads an unusually large volume of files from SharePoint, deviating from their historical access pattern.
\item \textbf{Unusual Login Times:} A user in Teams logs in at 3 AM despite never accessing the platform outside business hours.
\item \textbf{Abnormal File Modifications:} A script or bot suddenly renames and moves thousands of files in Google Drive, resembling a ransomware attack.

\end{itemize}

Each cloud platform has dedicated security policies based on its architecture and use cases as mention in Table \ref{tab:platform_anomalies}

\begin{table}[H]
    \centering
    \caption{Cloud Platform Behavior-Based Anomaly Policies}
    \label{tab:platform_anomalies}
    \begin{tabular}{>{\raggedright\arraybackslash}p{3.5cm} 
                    >{\raggedright\arraybackslash}p{7.5cm} 
                    >{\raggedright\arraybackslash}p{4cm}}
        \toprule
        \textbf{Platform} & \textbf{Security Policy Trigger} & \textbf{Potential Risk} \\
        \midrule
        SharePoint & Abnormal number of files updated & Unauthorized bulk edits, ransomware \\
        OneDrive & Access from non-compliant devices & Data leaks, unauthorized access \\
        Google Drive & Mass file deletion event & Insider threat, accidental data loss \\
        Teams & Suspicious external file sharing & Data exfiltration \\
        Box & Bulk unauthorized downloads & Intellectual property theft \\
        \bottomrule
    \end{tabular}
\end{table}

\end{enumerate}

\subsubsection{ AI-Driven Security Recommendations
}
The system integrates an on-premises AI model powered by DeepSeek Large Language Models (LLMs) to enhance security operations. Unlike traditional static alerts, this AI provides \textbf{context-aware security recommendations} based on real-time events \textbf{without triggering automated remediation actions}~\cite{wrixte2024context}.

\begin{enumerate}
    \item \textbf{Context-Aware Risk Analysis}

    Instead of relying solely on predefined thresholds, the AI analyzes contextual factors to generate actionable insights for security teams. It evaluates:
    \begin{itemize}
        \item User’s historical behavior – Has the user previously performed similar actions, such as large-scale data downloads?
        \item Organizational norms – Is this behavior typical for their department or role?
        \item Environmental context – Did the activity originate from a recognized, trusted device or network?
    \end{itemize}

    By considering these parameters, the \textbf{AI generates an informed security recommendation}, helping analysts understand the risk \textbf{in context} rather than reacting to isolated events~\cite{tcs2025context}.

    \item \textbf{AI-Generated Security Recommendations}

    Our system leverages an on-premises \textbf{DeepSeek LLM to provide real-time security insights without taking direct remediation actions}. When a high-risk score is assigned, a policy is violated, or a high-risk activity occurs, the AI dynamically analyzes the event and generates contextual recommendations.

    Instead of relying on static thresholds, \textbf{the AI follows a chain-of-thought reasoning} process to assess:

    \item \textbf{Why On-Prem AI?}

    The AI model runs entirely on-premises to maintain data sovereignty and privacy, ensuring no sensitive information leaves the system~\cite{kinoshita2025deepseek}. This closed-loop security approach prevents exposure to external cloud services while enabling real-time, context-rich security recommendations.

    The AI follows a chain-of-thought reasoning process, analyzing past activity patterns, organizational context, and real-time security events to provide accurate, actionable insights. This empowers security teams to make informed decisions rather than rely on rigid automation.

\end{enumerate}

\subsubsection{ Visual Analytics and Risk-Based Dashboards
}

Following \textbf{Policy-Based Risk Analysis \& Response Automation}, the system transitions into a visual analytics phase. This phase extracts and presents key security insights in a structured and interactive format to facilitate rapid decision-making. The \textbf{risk scoring system} fuels these visualizations and drives the intuitive dashboards, alerts, and graphs.
\begin{enumerate}

\item {Urgent Tab}
This section prioritizes \textbf{high-risk incidents} and supports rapid triage across multiple dimensions:
\begin{itemize}
    \item \textbf{Campaign Level:} Identifies security threats associated with specific risk campaigns.
    \item \textbf{User Level:} Highlights high-risk users needing immediate investigation.
    \item \textbf{Data Level:} Flags sensitive data breaches or policy violations.
    \item \textbf{Application Level:} Monitors anomalies within specific enterprise applications.
\end{itemize}

\item {Overview Tab}
Provides a \textbf{holistic view of system security}, summarizing high-level risk metrics:
\begin{itemize}
    \item \textbf{User Insights:} Categorizes users based on behavioral and contextual risk scores.
    \item \textbf{Campaign Insights:} Visualizes ongoing and completed risk detection campaigns.
    \item \textbf{Alert Distribution:} Displays alerts categorized by severity.
    \item \textbf{Risk Factors:} Summarizes data sensitivity, cross-platform vulnerabilities, and recent security trends.
\end{itemize}

\item {Analytics Page}
Offers \textbf{real-time security insights} through advanced data visualizations:
\begin{itemize}
    \item \textbf{Incident \& Risk Activities Graph:} Plots risk-based activity trends over time.
    \item \textbf{Risk Activity Analysis:} Detects behavioral anomalies and potential insider threats.
    \item \textbf{Data Breach Prevention Insights:} Highlights data leak vectors and sensitive information flow.
\end{itemize}

\item {Campaign Page}
Enables creation, tracking, and evaluation of risk mitigation campaigns:
\begin{itemize}
    \item \textbf{Overall Campaign Performance:} Measures detection efficacy and resolution time.
    \item \textbf{Individual Campaign Insights:} Tracks user engagement and policy violations per campaign.
\end{itemize}

\item {Users Page}
Delivers \textbf{user-specific risk intelligence} across individual and organizational levels:
\begin{itemize}
    \item Detailed user profiles, risk scores, and behavioral history.
    \item Segmented views for user grouping and role-based investigation.
\end{itemize}

\item {Notification Page}
Aggregates \textbf{real-time alerts} to maintain a proactive security posture:
\begin{itemize}
    \item Displays ongoing incidents and policy violations.
    \item Enables security teams to respond and resolve threats swiftly.
\end{itemize}

\vspace{0.5em}
The entire visualization framework is \textbf{powered by the risk scoring system}, ensuring continuous, data-driven insights and enhancing threat detection and response capabilities.

\end{enumerate}

\section{\textbf{\textbf{Evaluation and Results}}}
Our system's evaluation demonstrates significant improvements in risk detection accuracy, response efficiency, and overall security management. The results include key performance metrics, graphical representations, and comparative analysis.

\subsection{PRISM – Privilege-based Risk \& Insider Scoring Mechanism}

\begin{figure}[H]
    \centering
    \includegraphics[width=1\linewidth]{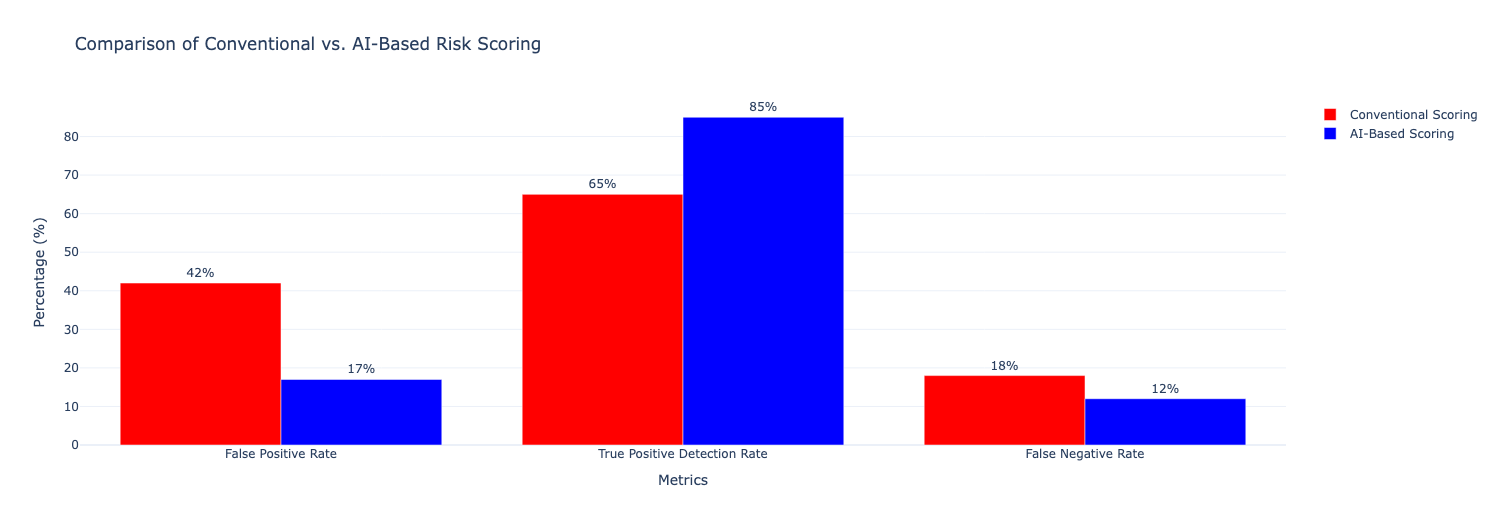}
    \caption{False Positive Rate in PRISM vs. AI-Based Risk Scoring}
    \label{fig:prism_ai}
\end{figure}

Our initial risk-scoring model assessed security risks using predefined rules and static thresholds. While effective at detecting common threats, it had notable limitations, including a high false-positive rate and a lack of adaptability to evolving risk patterns.
As shown in Figure \ref{fig:prism_ai}, our AI-based risk-scoring model significantly outperforms the PRISM approach. The false positive rate has been reduced from 42\% to 17\%, leading to 2.5x fewer false alerts. The positive detection rate has increased from 65\% to 85\%, making the model 1.3x more effective at identifying real threats. The false negative rate has also dropped from 18\% to 12\%, reducing missed risks by 1.5x.
The table \ref{tab:ai_vs_prism} provides a direct comparison of PRISM vs. AI-based scoring, along with ratio-based improvements:

\begin{table}[H]
\centering
\caption{Performance Comparison: PRISM Scoring vs. AI-Based Scoring}
\label{tab:ai_vs_prism}
\begin{tabular}{p{4.5cm} p{2.5cm} p{2.5cm} p{5.5cm}}
\toprule
\textbf{Metric} & \textbf{PRISM Scoring} & \textbf{AI-Based Scoring} & \textbf{Improvement / Ratio} \\
\midrule
False Positive Rate & 42\% & 17\% & 59\% reduction \newline 2.5x lower (AI reduces false positives) \\

True Positive Detection Rate & 65\% & 85\% & 30\% increase \newline 1.3x higher (AI detects more threats) \\
False Negative Rate & 18\% & 12\% & 33\% reduction \newline 1.5x lower (AI reduces missed risks) \\
\bottomrule
\end{tabular}
\end{table}

We trained our model with user feedback over 12 weeks to achieve these results, incorporating approximately 300 training instances in the initial three weeks. Our custom model was continuously evaluated on a dataset annotated by field experts and the administrator, ensuring high accuracy and relevance. Over time, the model kept improving, adapting to user feedback and administrator preferences.
\vspace{0.5em}

Furthermore, the AI model continuously improves with additional training data, adapting to organization-specific risk patterns. Over time, it will learn from administrator preferences, allowing for a customized risk-scoring approach that aligns with the organization's unique security needs.

\subsection{Accuracy of AI-Based Risk Scoring Model}
\begin{figure}[H]
    \centering
    \includegraphics[width=0.7\linewidth]{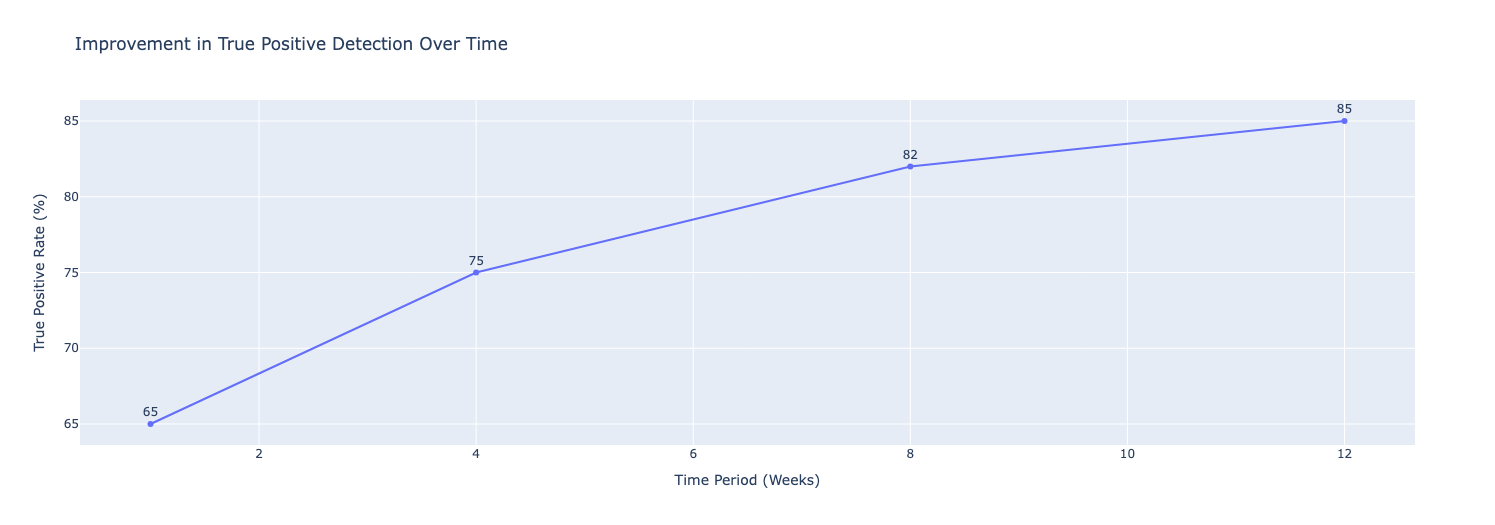}
    \caption{\textit{\textbf{Improvement in True Positive Detection Over Time}}}
    \label{fig:true_poitive}
\end{figure}

Risk detection improved significantly by introducing our AI-powered risk scoring model powered by an autoencoder neural network. After continuous learning from user feedback, the model adjusted dynamically, achieving a 17\% false positive rate—less than half of the rule-based system. See figure\ref{fig:true_poitive} and Table \ref{tab:positive_rate_progression}.

\begin{table}[H]
\centering
\caption{Improvement in Actual Positive Rate Over Time}
\label{tab:positive_rate_progression}
\begin{tabular}{|l|c|}
\hline
\textbf{Period (Weeks)} & \textbf{Actual Positive Rate (\%)} \\
\hline
Week 1 & 65\% \\
\hline
Week 4 & 75\% \\
\hline
Week 8 & 82\% \\
\hline
Week 12 & 85\% \\
\hline
\end{tabular}
\end{table}

\subsection{User Feedback \& Model Adaptation}

Over a 60-day testing period, security analysts manually reviewed and adjusted 12\% of AI-generated risk scores to fine-tune the model’s accuracy. This iterative feedback process led to a significant improvement in risk classification.

\vspace{0.5em}

\begin{figure}[H]
    \centering
    \includegraphics[width=0.75\linewidth]{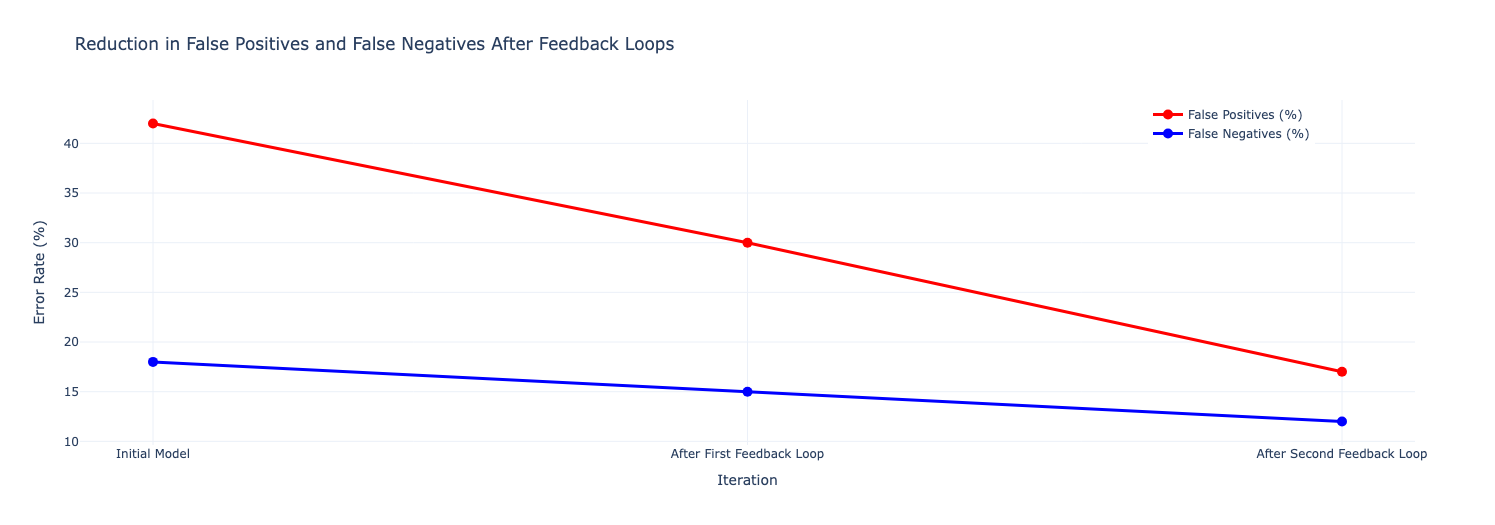}
    \caption{ Reduction in False Positives After User Feedback Loops}
    \label{fig:Redu_false}
\end{figure}

As shown in Figure \ref{fig:Redu_false} and the accompanying table\ref{tab:feedback_loops}, the initial AI model had a false positive rate of 42\% and a false negative rate of 18\%. After the first feedback loop, where analyst corrections were incorporated into retraining, false positives dropped to 30\% and false negatives to 15\%. Following a second feedback loop, the model further improved, reducing false positives to 17\% and false negatives to 12\%.

\vspace{0.5em}

This demonstrates the effectiveness of continuous learning—each iteration refines the AI’s decision-making, reducing unnecessary alerts while improving real threat detection. As more feedback is incorporated, the model adapts dynamically to organizational risk patterns, ensuring a more accurate and customized risk-scoring system.

\begin{table}[H]
\centering
\caption{Model Improvement Through Feedback Loops}
\label{tab:feedback_loops}
\begin{tabular}{|l|c|c|}
\hline
\textbf{Iteration} & \textbf{False Positives (\%)} & \textbf{False Negatives (\%)} \\
\hline
Initial Model & 42\% & 18\% \\
\hline
After First Feedback Loop & 30\% & 15\% \\
\hline
After Second Feedback Loop & 17\% & 12\% \\
\hline
\end{tabular}
\end{table}

\subsection{Policy Violation Identification}

During testing, our policy-based risk analysis engine detected 78 critical violations over a month, covering various security threats such as \textbf{unauthorized privilege escalations, data handling breaches, and non-compliant device connections.}

\begin{figure}[H]
    \centering
    \includegraphics[width=1\linewidth]{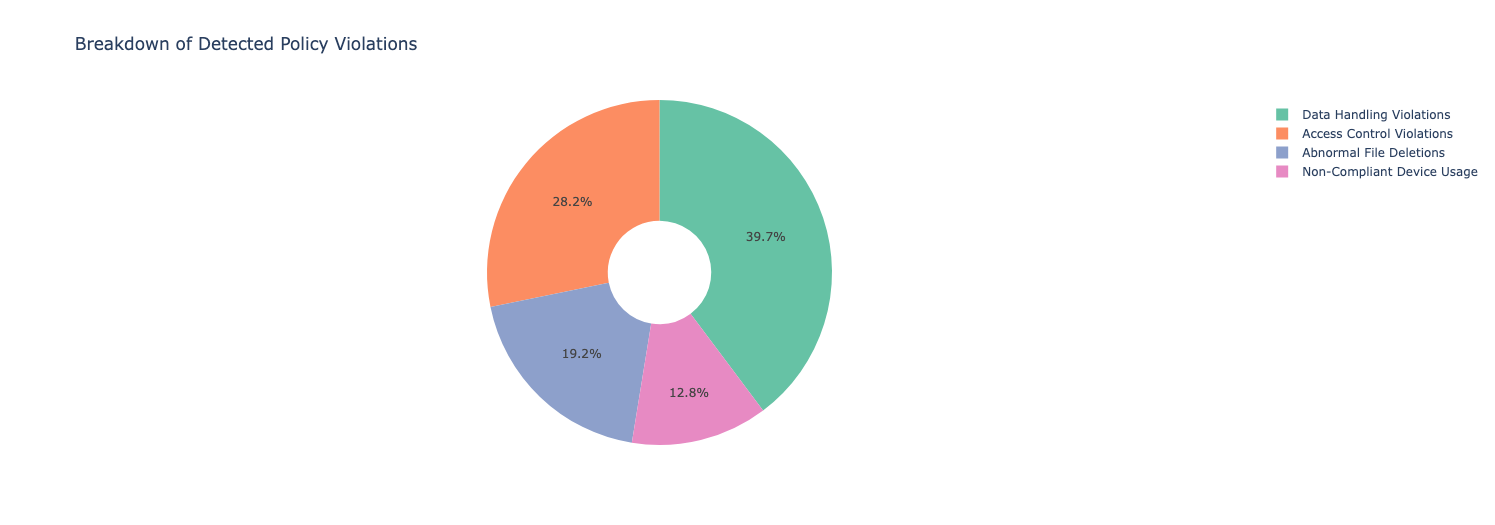}
    \caption{Breakdown of Detected Policy Violations}
    \label{fig:enter-label}
\end{figure}

As shown in \textbf{Figure \ref{fig:enter-label}} and the accompanying table\ref{tab:policy_actions}, the highest number of violations were in data handling, with 31 detected cases, followed by access control violations (22 cases). In response to these risks, automated security measures were applied, including revoking privileges (15 cases), restricting file access (20 cases), and flagging suspicious user activity (10 cases).

The \textbf{policy enforcement system }continuously monitors security events and applies real-time mitigation actions, reducing manual intervention and improving overall compliance. 

\begin{table}[H]
\centering
\caption{Policy Violations and Automated Actions}
\label{tab:policy_actions}
\begin{tabular}{p{4.5cm} p{2cm} p{6cm}}
\toprule
\textbf{Policy Category} & \textbf{Violations Detected} & \textbf{Automated Actions Taken} \\
\midrule
Access Control Violations & 22 & 15 (Privilege Revoked) \\
Data Handling Violations & 31 & 20 (File Access Restricted) \\
Abnormal File Deletions & 15 & 10 (User Flagged) \\
Non-Compliant Device Usage & 10 & 7 (Device Disconnected) \\
\bottomrule
\end{tabular}
\end{table}

\subsection{Performance \& Scalability}

The system demonstrated exceptional performance under high log ingestion loads, efficiently processing up to \textbf{10 million log events daily} while maintaining a \textbf{sub-300ms query response time}. Performance tests across different ingestion rates, as shown in \textbf{Table 1}, revealed that at \textbf{1,000 logs per second}, the system responded within \textbf{120ms}, while at \textbf{10,000 logs per second}, response time increased slightly to \textbf{190ms}. Even under an extreme load of 100,000 logs per second, the system maintained a \textbf{250ms} response time, ensuring real-time threat detection and analysis.

\begin{table}[H]
\centering
\caption{System Performance: Log Ingestion vs. Query Response Time}
\label{tab:log_performance}
\begin{tabular}{p{5cm} p{5cm}}
\toprule
\textbf{Log Ingestion Rate (Logs/sec)} & \textbf{Query Response Time (ms)} \\
\midrule
1,000     & 120 \\
10,000    & 190 \\
100,000   & 250 \\
\bottomrule
\end{tabular}
\end{table}

These results indicate the system’s \textbf{high scalability}, making it suitable for organizations of all sizes, from SMBs to large enterprises with vast security logs. The \textbf{low-latency query execution} enables security analysts to retrieve insights instantly, facilitating \textbf{faster threat detection and mitigation}. Additionally, as seen in \textbf{Table \ref{tab:log_performance}}, the system maintains optimal performance even as log ingestion rates increase, proving its \textbf{efficiency under heavy workloads}. The architecture also supports future scalability, as further optimizations, such as \textbf{advanced indexing strategies and parallel processing}, can enhance performance even under higher loads. This ensures that security teams can respond promptly to emerging threats, improving overall \textbf{incident response time and cyber risk management.}

\subsection{Comprehensive Impact of Insider Risk Management (IRM) System}

\begin{figure}[H]
    \centering
    \includegraphics[width=1.1\linewidth]{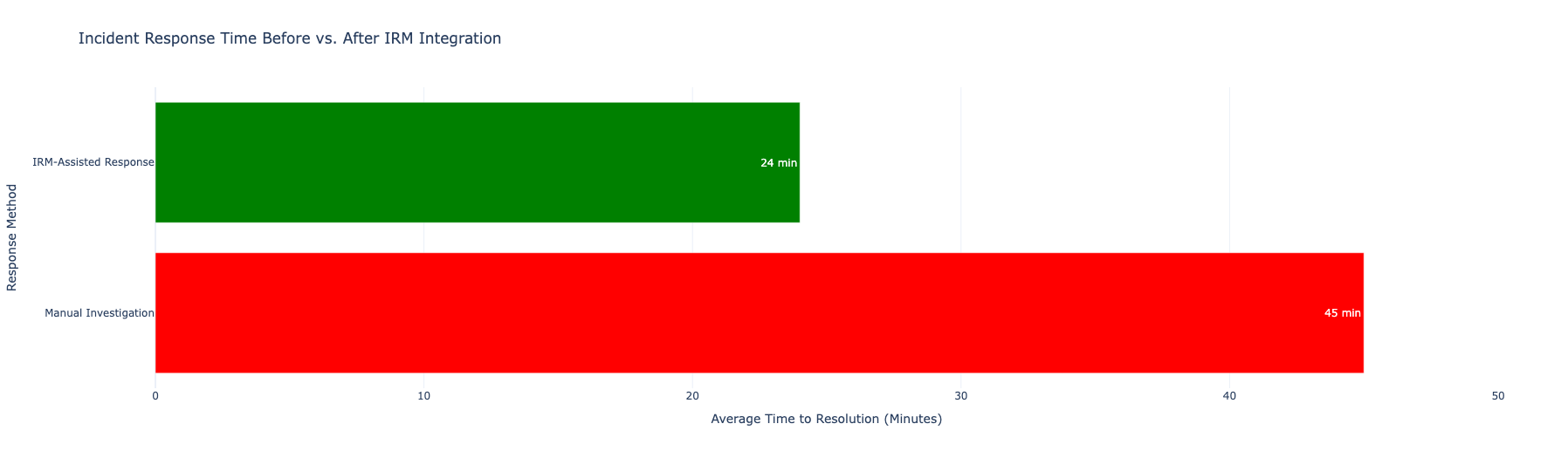}
    \caption{Incident Response Time Before vs. After IRM Integration}
    \label{fig:After_IRM_Integration}
\end{figure}

Implementing the Insider Risk Management (IRM) system has significantly strengthened security operations by integrating AI-driven risk scoring, policy violation detection, and sensitive data classification. One of the most impactful outcomes has been a 47\% reduction in incident response time, primarily driven by automated risk assessments, real-time policy enforcement, and AI-assisted decision-making. The system enhances efficiency through automated policy violation detection, instantly flagging unauthorized actions and minimizing investigation delays. Additionally, risk-based prioritization ensures that security teams focus on the most critical threats first, optimizing resource allocation. Integrating sensitive data classification allows for context-aware alerts, improving the accuracy of risk assessments and reducing false positives. Furthermore, a streamlined investigation workflow provides security teams with pre-analyzed insights, reducing the manual effort required to correlate security events.
These enhancements have led to a significant improvement in incident response efficiency, as demonstrated in Table \ref{tab:irm_efficiency}. The average resolution time decreased from 45 minutes (manual investigation) to 24 minutes with IRM-assisted response. This reduction is further visualized in Figure \ref{fig:After_IRM_Integration}, which presents a comparative bar chart highlighting the efficiency gains achieved through AI-driven automation.

\begin{table}[H]
\centering
\caption{Incident Response Efficiency – Before vs. After IRM Implementation}
\label{tab:irm_efficiency}
\begin{tabular}{p{6cm} p{5cm}}
\toprule
\textbf{Response Method} & \textbf{Average Time to Resolution} \\
\midrule
Manual Investigation     & 45 minutes \\
IRM-Assisted Response    & 24 minutes \\
\bottomrule
\end{tabular}
\end{table}

\section{\textbf{Conclusion and Future Work}}

This research has demonstrated the significant potential of AI-powered Insider Risk Management (IRM) systems to enhance organizational security by leveraging advanced machine learning models, behavioral analytics, and real-time data monitoring. The proposed framework strengthens the detection accuracy and response speed by fusing diverse data sources and performing contextual risk analysis, effectively reducing the exposure window to insider threats. Major contributions include the development of an AI-driven risk assessment model, deployment of behavioral baselining techniques, and validation of the system’s effectiveness in operational environments, leading to reduced false positives, improved incident response times, and scalable deployment across both on-premises and cloud infrastructures. Moving forward, several enhancements are envisioned: implementing federated learning for privacy-preserving AI training, integrating explainable AI techniques for improved transparency, aligning the IRM system with Zero Trust security frameworks, enriching behavioral anomaly detection through graph-based analysis, enabling real-time response mechanisms, expanding cross-platform compatibility for hybrid environments, and embedding automated compliance audits to adapt to evolving regulations. By pursuing these directions, the IRM system can become an even more powerful, adaptable, and comprehensive solution, equipping organizations with the necessary tools to proactively mitigate emerging insider threats and protect critical assets.

\bibliographystyle{unsrt}  


\end{document}